\documentstyle[aas2pp4]{article}

\def\mm{M\'endez }

\begin{document}

\title{Discovery of a Second Kilohertz QPO in the X--ray Binary 4U~1735--44}

\author{Eric C. Ford\altaffilmark{1}, Michiel van der
Klis\altaffilmark{1}, Jan van Paradijs\altaffilmark{1,2}, 
Mariano \mm\altaffilmark{1,3}, Rudy Wijnands\altaffilmark{1} 
and Philip Kaaret\altaffilmark{4}}

\authoremail{ecford@astro.uva.nl}
\altaffiltext{1}{Astronomical Institute, ``Anton Pannekoek'',
University of Amsterdam, Kruislaan 403, 1098 SJ Amsterdam,
The Netherlands}
\altaffiltext{2}{University of Alabama in Huntsville, Department
of Physics, Huntsville, AL 35899}
\altaffiltext{3}{Facultad de Ciencias Astron\'omicas y Geof\'{\i}sicas,
Universidad Nacional de La Plata, Paseo del Bosque S/N, 1900 La Plata,
Argentina.}
\altaffiltext{4}{Harvard--Smithsonian Center for Astrophysics, 
60 Garden Street, Cambridge, MA 02139}

~~

\centerline{\bf {Accepted by ApJL: 30 September 1998}}

\begin{abstract}

In recent observations with the Rossi X--Ray Timing Explorer we have
detected two simultaneous quasi-periodic oscillation (QPO) peaks in
the low mass X--ray binary and atoll source 4U 1735--44.  The lower
and higher frequency QPOs have frequencies varying between 632 and 729
Hz, and 982 and 1026 Hz, respectively. The fractional rms amplitudes
are 3.7 to 8.1\% and 5.0 to 5.8\%. The frequency separation between
the two QPOs changes from $341\pm7$ Hz to $296\pm12$ Hz.  The inferred
mass accretion rate during our observations is relatively low compared
to that during the previous observations, where only a single QPO was
present. There is weak evidence that the frequency of the QPOs
correlates with the mass accretion rate, as observed in other
binaries.  Five X--ray bursts were recorded with no detectable
oscillations with upper limits for the rms fraction of 4\% to 13\%.

\end{abstract}


\section{Introduction}

Observations with the Rossi X--Ray Timing Explorer (RXTE) have shown
that many low--mass X--ray binaries with neutron stars exhibit
oscillations in their X--ray flux at frequencies near 1 kHz (for
reviews and references see van der Klis 1998, Swank 1998).  It is not
clear how the kilohertz QPOs are produced, but they most likely
originate very close to the neutron star.

Typically two QPO peaks are simultaneously observable in the the
persistent X--ray emission. Early data suggested that only a single
QPO was present in some binaries. Further observations, however,
revealed a second QPO peak in most cases; an example is 4U~1608--52
(Berger et al. 1996, \mm et al. 1998c). The only current exceptions
are 4U~1735--44 (Wijnands et al. 1998) and Aql~X-1 (Zhang et
al. 1998a) which have shown only one QPO. Here we show that
4U~1735--44 in fact exhibits two QPO peaks.

In addition to the double QPOs present in the persistent emission of
X--ray binaries there is one nearly coherent oscillation seen during
some X--ray bursts.  The presence of three distinct QPOs in some of
these X--ray binaries offers the most intriguing clue yet to the
physical mechanisms at work.  In four cases the difference in
frequency between the persistent QPOs is close to the frequency of the
oscillation during the bursts or close to half that value. The simple
interpretation is that the burst oscillation is a modulation from the
spin of the neutron star, the highest frequency QPO represents a
Keplerian frequency, and a beat--frequency mechanism (Alpar \& Shaham
1985, Strohmayer et al. 1996) accounts for the lowest frequency QPO.

This interpretation has recently been called into question by new
data.  The frequency difference of the double QPOs is actually not
constant in both the Z-source Sco~X-1 (van der Klis et al. 1997) and
the atoll source 4U~1608--52 (\mm et al. 1998a). Furthermore, in
4U~1636--53, the frequency of the burst oscillations does not exactly
match the frequency difference (\mm et al. 1998b). This casts doubt on
the beat-frequency mechanism or at least complicates such a
model. Further observations of systems with double QPOs and X--ray
bursts are clearly desirable.

The RXTE observations of 4U~1735--44 in this paper were obtained by a
trigger designed to capture this source at a relatively low mass
accretion rate. In Section 2 we present our observations and
results. In Section 3 we discuss the QPO properties in the context of
other measurements.

\section{Observations \& Results}

In May 1998 we initiated observations of 4U~1735--44 with RXTE based
on a low flux measurement by the RXTE All-Sky Monitor. The
observations were conducted on 30 and 31 May 1998 and yielded roughly
21 ksec of usable data divided into intervals by the satellite
orbit. We generated Fourier power spectra from `event mode' data from
the proportional counter array (PCA) with a time resolution of 122
$\mu$sec. Except where noted, we use the entire PCA energy band, most
sensitive from 2 to 30 keV. For color and energy spectral analysis we
use the `Standard 2' mode data and eliminate the last of the five
detectors which is off during part of the 30 May observation. We
perform background subtraction with the version 2.0c of the RXTE
background estimator (Stark et al. 1998).

\begin{figure*}
\figurenum{1}
\epsscale{1.8}
\plotone{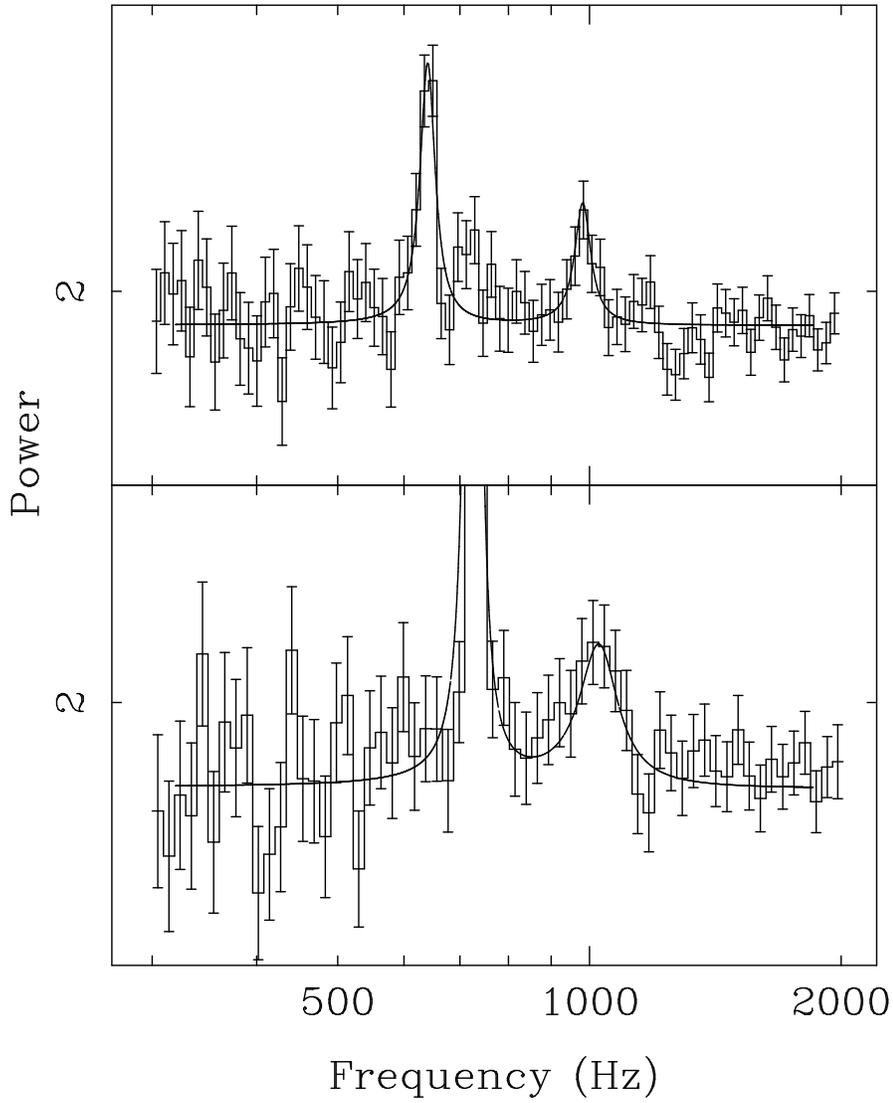} 
\caption{Power density spectrum for observations starting 30 May 1998
7:50:23 (top) and 31 May 11:01:19 UTC (bottom). The lower frequency
peak of 31 May is off--scale. See Table~1 for more
information.}
\label{fig:pds}
\end{figure*}

In the May 1998 observations we detect two simultaneous kilohertz
QPOs.  Figure~\ref{fig:pds} shows example Fourier power spectra from
two intervals.  In these spectra the significance of the two QPOs are
6.4$\sigma$ and 3.7$\sigma$ (30 May) and 17.8$\sigma$ and 4.1$\sigma$
(31 May). We report here only detections with a significance greater
than $3\sigma$. Complete results are summarized in
Table~\ref{tbl:fits}.  The double QPOs are detectable in three
separate time intervals. The difference in frequency between the two
QPOs has an average value of $326\pm6$ Hz. This values does however
change from $341\pm7$ to $296\pm12$ Hz. These values are different
with a significance of 3.1$\sigma$.

The X--ray color--color diagram is shown in Figure~\ref{fig:cd}.  For
completeness we include the observations from August to October 1996
(Wijnands et al. 1998). The colors and the properties of Fourier power
spectra change in a way that is well known for atoll sources (Hasinger
\& van der Klis 1989). In all the 1996 observations the source is in
the `banana state' of the atoll sources (Wijnands et al. 1998). The
power spectra show a power law noise component below 1 Hz. In the
leftmost part of the banana branch (observations from August 1996) the
`high frequency noise' component known in atoll sources is also
present with an rms fraction of 3.6\% (2--18 keV and 0.01--100 Hz).
The observations of 30 and 31 May 1998 sample the source in a
different state: the `island state'.  These data occupy a roughly
circular region in the color diagram. The power spectra can be
described by a broken power law, representing the high frequency noise
component. The rms fraction of this component is strongest here,
$6.1\pm0.3$\% to $6.6\pm0.2$\%.  These data represent the most extreme
island state yet observed in this source as judged by the strength of
the high frequency noise component. The strongest high frequency noise
previously observed was $4.0\pm0.4$\% for similar energies (Hasinger
\& van der Klis 1989).  We do not detect features similar to the 30 or
60 Hz features seen in the 1996 data (Wijnands et al. 1998).

There is evidence that in 4U~1735--44, like other binaries, the source
state and the frequencies of the QPOs are related. In the 1996
observations, in the lower part of the banana, there is a QPO at 1144
to 1161 Hz which is most likely the higher frequency QPO (Wijnands et
al. 1998). In the present data, an island state, the frequency of this
QPO is 982 to 1026 Hz. 

We have also analyzed the energy spectra. Fitting the spectra with an
absorbed blackbody plus power law components, we find a total absorbed
X--ray flux of $3.2-5.6\times10^{-9}$ erg~cm$^{-2}$~s$^{-1}$ (2--20
keV). The lowest fluxes are from the island states in May 1998, while
the highest fluxes are in the banana state from 1996. The
corresponding luminosity is $3.2-5.6\times10^{37}$ erg~s$^{-1}$ for a
distance of 9.2 kpc (Van Paradijs, Penninx \& Lewin 1988), making 4U
1735--44 one of the most luminous atoll sources to show kilohertz QPOs.

There are five X--ray bursts in the 30 May 1998 observations.  We
detect no oscillations in any of the bursts, searching Fourier power
spectra from both the full PCA energy band and the 2 to 5 keV band. We
find upper limits of 4\% to 7\% for the full band (the better limits
being at the peak of the burst) and 8\% to 13\% for the 2 to 5 keV
band. These limits are much smaller than the amplitudes of bursts
oscillations observed in other sources, but similar to the upper
limits in bursts where there are no oscillations (Strohmayer, Zhang \&
Swank 1997).

\section{Discussion}

We have discovered a second kilohertz QPO in 4U 1735--44 and observed
a link between the QPO properties and the inferred mass accretion
rate.  4U~1735--44 was one of only two sources in which only a single
QPO had been securely detected. The separation in frequency between
the two QPOs decreases with increasing inferred mass accretion
rate. The change has a 3.1$\sigma$ significance. The average frequency
separation is $326\pm6$ Hz, similar to the value observed in other
low-mass X--ray binaries.  Under the simple beat--frequency
interpretation of the QPOs the frequency difference would equal to the
spin frequency of the neutron star. This interpretation, however, is
called into question by an observed change in the frequency
separation. The frequency separation also changes in Sco X-1 (van der
Klis et al. 1997) and 4U~1608-52 (\mm et al. 1998a), becoming smaller
at higher accretion rates.  If any burst oscillations are detected in
future observations, the beat frequency prediction can be
tested. There is strong motivation for such measurements given the
recently noted discrepancy in these frequencies in 4U~1636--53 (\mm et
al. 1998b).

\begin{figure*}
\figurenum{2}
\epsscale{1.8}
\plotone{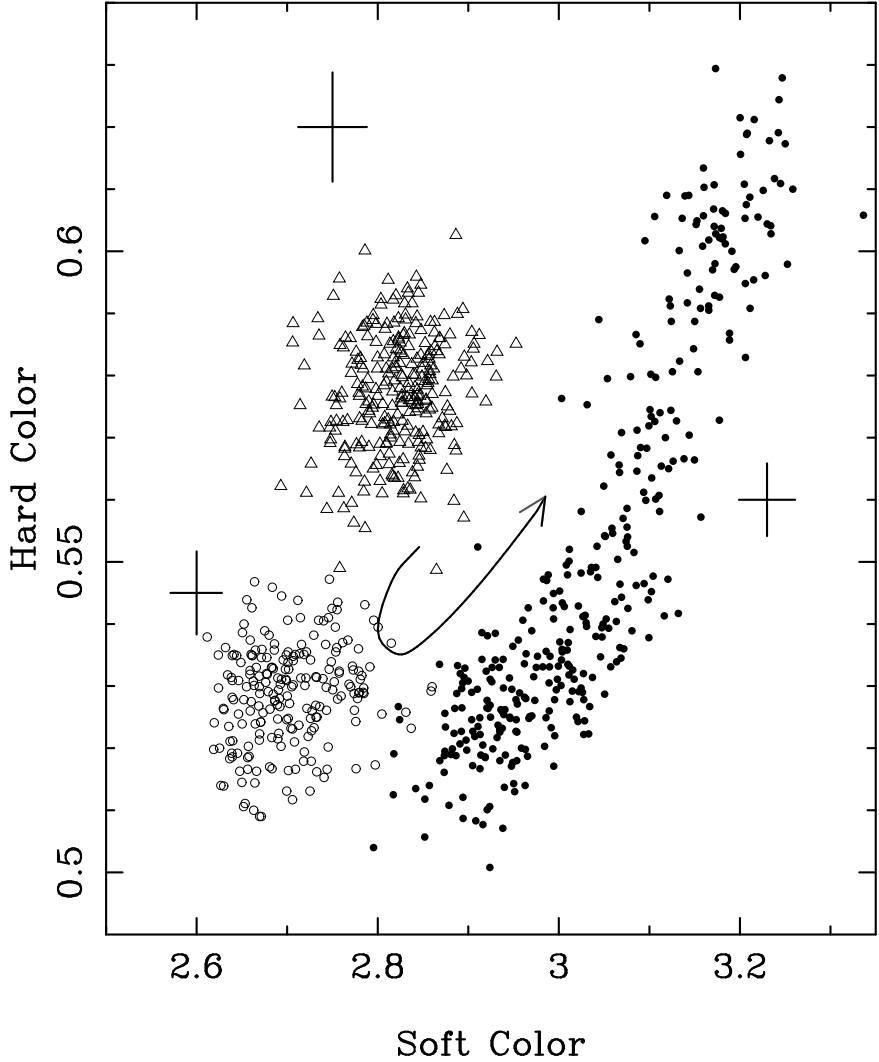} 
\caption{X--ray color-color diagram of 4U 1735--44. The data symbols
are: May 1998 (triangles), 1 August 1996 and 28 September 1996 (open
circles), and 1 and 4 September 1996 and 29 October 1996 (filled
circles). The arrow shows the direction of increasing inferred mass
accretion rate. Colors are defined by ratios of PCA count rates in the
bands 3.4--6.3/2.0--3.4 keV (soft color) and 9.6--15.7/6.3--9.6 keV
(hard color). Data points are 48 second time bins from the first four
detector units with background subtraction.  Typical errors are
shown.}
\label{fig:cd}
\end{figure*}

The present data also show a link between the properties of the QPOs
and the the inferred mass accretion rate. This is illustrated most
clearly in Figure~\ref{fig:cd}. The mass accretion rate likely
increases going from the `island' to the `banana' states as shown by
the arrow.  The QPOs are present only at the relatively lower mass
accretion rates.  At the lowest accretion rates there are two QPOs,
while at slightly higher rates there is one QPO. At the highest rates
there are none.

This fits into a general pattern seen in atoll sources. 4U~1820--30
has a similar color diagram and the kilohertz QPOs were apparently
detected only at intermediate accretion rates (Zhang et al. 1998b).
During a recent outburst of 4U~1608--52 the QPOs were detected at the
intermediate rates, at the lower end of the banana branch and the
island states closet to the banana (\mm et al. 1998d). Similarly in 4U
1705--440, QPOs are seen only at intermediate accretion rates (Ford,
van der Klis \& Kaaret 1998). At the highest accretion rates, farthest
into the banana branches, the QPOs are absent as seen in numerous
sources: 4U 1636--53 (Zhang et al. 1996; Wijnands et al. 1997),
4U~1820--30 (Smale, Zhang \& White 1997, Zhang et al. 1998b), KS
1731--260 (Wijnands \& van der Klis 1997), and 4U 1705--440 (Ford, van
der Klis \& Kaaret 1998). In these high states the upper limits are
quite strong (typically 2 to 4\% rms). In the island states, which
have lower count rates, the upper limits are less stringent.  On
certain occasions QPOs have been detected in island states in the
following sources: 4U~0614+091 (\mm et al. 1997), 4U~1608--52 (Yu et
al. 1997, \mm et al. 1998d) and 4U~1728--34 (Ford \& van der Klis
1998, Strohmayer et al. 1996).

The mass accretion rate is linked not only to the presence of the
QPOs but to their frequencies as well. In the present data the
frequency of the QPO is highest in the banana state.  The data are
insufficient however to quantitatively distinguish a correlation of
the QPO frequency with the properties of the noise at lower
frequencies. Such a trend is seen in other atoll sources, most
notably 4U~1728--34 (Ford \& van der Klis 1998).  This correlation can
be explain by current models for the production of kilohertz QPO
(e.g. Miller, Lamb \& Psaltis 1997; Titarchuk, Lapidus \& Muslimov
1998).

Observations therefore point to a clear link between the presence and
properties of the QPOs and the apparent mass accretion rate of a given
source. There is, however, no such correlation between sources. For
example 4U 1735--44, 4U 1608--52 and 4U 0614+091 are each separated by
about a decade in luminosity, nevertheless the QPO properties are
nearly identical. This is the most enigmatic observational fact about
kilohertz QPOs.

ECF acknowledges support by the Netherlands Foundation for Research in
Astronomy with financial aid from the Netherlands Organization for
Scientific Research (NWO) under contract numbers 782-376-011 and
781-76-017.  M. M. is a fellow of the Consejo Nacional de
Investigaciones Cient\'{\i}ficas y T\'ecnicas de la Rep\'ublica
Argentina. JvP acknowledges support from NASA grants NAG5-4482,
NAG5-7382 and NAG5-7415. PK acknowledges support from NASA grants
NAG5-7405 and NAG5-7407.


\begin{deluxetable}{lcccccc}
\tablenum{1}
\tablewidth{40pc}
\tablecaption{4U~1735--44 May 1998 Observations}
 
\tablehead{ 
\colhead{Observation} & \colhead{$T$} & \colhead{$R$} &
\colhead{Frequency} & \colhead{$\Delta \nu$} & \colhead{FWHM} & 
\colhead{rms} \nl
\colhead{(1998)} & \colhead{(sec)} & \colhead{(counts s$^{-1}$)} & 
\colhead{(Hz)} & \colhead{(Hz)} & \colhead{(Hz)} & \colhead{(\%)} \nl
}

\startdata
30 May  7:50:23 & 3506 & 1308 & $981.7\pm6.7$  & $341.2\pm7.2$ &
  $32\pm7$   &  $5.3\pm0.7$  \nl 
                &      &      & $640.5\pm2.5$  &               &
  $53\pm20$  &  $6.2\pm0.5$  \nl
30 May  9:25:20 & 1351 & 1282 & $631.9\pm6.3$  & & $38\pm20$ & $5.9\pm0.7$  \nl
30 May  9:49:38 & 1710 & 1299 & $633.3\pm5.7$  & & $39\pm14$ & $5.4\pm0.8$  \nl
30 May 11:01:19 & 3740 & 1358 & $709.5\pm0.4$  & & $23\pm2$  & $7.5\pm0.3$  \nl
30 May 12:38:33 & 2983 & 1376 & $1025.0\pm12.3$ & $296.4\pm12.3$ & 
  $84\pm47$ & $5.0\pm1.0$ \nl
                &      &      & $728.6\pm0.3$  & & $9\pm2$   & $7.2\pm0.2$  \nl
31 May 11:01:19 & 3767 & 1376 & $1025.7\pm17.0$ & $300.2\pm17.0$ &
  $135\pm43$ & $5.8\pm0.8$  \nl
                &      &      & $725.5\pm0.8$  & & $24\pm2$  & $8.1\pm0.3$  \nl
31 May 12:37:19 & 3807 & 1364 & $716.9\pm0.6$  & & $21\pm1$  & $3.7\pm0.9$  \nl

\tablecomments{ The start time of each observation is in UTC
(Universal Time, Coordinated). $T$ is the duration of the observation.
$R$ is the count rate background subtracted over the full PCA energy
band for 5 PCUs. The QPO frequencies, FWHM and rms fraction are listed as
calculated using data of the entire PCA energy band. $\Delta \nu$ is
the frequency difference between double QPOs. Errors are statistical
using $\Delta \chi^2 = 1$.}

\enddata
\label{tbl:fits}
\end{deluxetable}


\end{document}